# Irreversibility Field and Flux Pinning in MgB$_2$ with and Without SiC Additions

M.D. Sumption[1], M. Bhatia[1], E.W. Collings[1], M. Rhindfliesh[2],
M. Tomsic[2], Y. Hascicek[3], and S.X. Dou[4]

[1] LASM, Materials Science and Engineering Department,
OSU, Columbus, OH 43210, USA
[2] Hyper Tech Research, Inc. Columbus, OH 43210, USA
[3] The National High Magnetic Field Laboratory, Tallahassee FL, USA
ISEM, The University of Wollongong, Australia

**Abstract**

Critical current density was measured at 4.2 K for MgB$_2$ strands with and without SiC additions. In some cases measurements were performed on longer (1 m) samples wound on barrels, and these were compared to magnetic measurements. Most measurements were performed on short samples at higher fields (up to 18 T). It was found that in-situ processed strands with 10% SiC additions HT at 700-800°C show improved $H_r$ and $F_p$ values as compared to control samples, with $H_r$ increasing by 1.5 T. At 900°C even larger improvements are seen, with $H_r$ reaching 18 T and $F_p$ values maximizing at 20 GN/m$^3$.

## Introduction

Many groups now fabricate MgB$_2$ wires [1-13], and PIT strands are very promising for current carrying applications. Typical present day strands incorporate Fe or Cu, with perhaps Cu-Ni or monel as an outer sheath. There are two main variants for PIT MgB$_2$ fabrication, ex-situ [1-4], and in-situ [6,10-13]. Each of these choices has some advantages and some disadvantages, we focus here on in-situ powders, aiming to look at the influence of SiC on $H_r$ and flux pinning, and to differentiate its influences in these two areas. Below we give a short background on MgB$_2$ strands, with an emphasis on $H_r$ to put the present efforts in context.

## Background

Grasso [1] using an *ex-situ* powder method to make MgB$_2$ tapes, found an irreversibility field, H$_r$, (the field at which the bulk current density goes to zero) just slightly higher that 12 T for field perpendicular to the tape face. An anisotropy factor of about 1.4 in $H_r$ has been reported for tapes (with the parallel orientation being higher than the perpendicular one [2]), indicating that rolling partially aligns the Mg and B planes parallel to the broad face of the tapes. Suo and Flukiger, using the ex-situ technique $H_{c2}$s of 11.9 and 15.1 T at 4 K for $H_\perp$ and $H_{//}$ (with a tape anisotropy of 1.3) [3]. Irreversibility fields were 8 and 10.4 T were measured at 4 K for $H_\perp$ and $H_{//}$, respectively [4]. Ball milling of MgB$_2$, also extensively employed by Flukiger [4] was seen to increase $H_r$ and $J_c$.

Matsumoto and Kumakura [6] investigated the use of $SiO_2$ and SiC in the *in-situ process*, finding that they were effective in improving $J_c$ considerably, as well as enhancing $H_r$ substantially. In this particular case, $J_c$ was sometimes increased by an order of magnitude with additions to make $J_c$ values of 1600-6500 A/cm$^2$ at 12 T, 4 K, with $H_r$ increasing from about 17 T to about 23 T at 4 K. Matsumoto and Kumakura [7] also tried $ZrSi_2$, $ZrB_2$ and $WSi_2$ additions to good effect. In a series of important papers Dou et al showed that SiC can significantly improve the properties of wires [14-16], although whether this is by improvements in pinning, or enhancements in the upper critical field, $H_{c2}$, as has been seen for thin films [18,19] is still not certain. This is the focus of the present work.

**The Upper Critical Field, $H_{c2}$, and its Enhancement through Impurity Scattering**

*Upper Critical Fields in Conventional Single-Band Superconductors*

For dirty non-paramagnetically limited single-band low temperature superconductors a well known formula for $H_{c2}(0)$ at zero temperature can be obtained by starting with Maki's dirty-limit $Hc_{20}$ and inserting an expression for the BCS zero-K thermodynamic critical field, $H_{c0}$, and find [20]

$$H_{c20} = 3.06 \times 10^4 \, \gamma \rho T_c \qquad (1)$$

where $\rho$ is the impurity resistivity and $\gamma$ the electronic specific heat coefficient. Alternatively we can predict a value for $H_{c20}$ from the slope of $H_{c2}$ at $T_c$. This time we begin with a Maki dirty-limit nonparamagnetic theory but at $T = T_c$. We combine the Maki $(dH_{c2}/dT)_{Tc}$ with the BCS $(dH_c/dT)_{Tc}$ [20] and find first of all that

$$\left(\frac{dH_{c2}}{dT}\right)_{Tc} = 4.49 \times 10^4 \, \rho \gamma \qquad (2)$$

Finally by combining (2) with (1) we obtain the frequently quoted expression:

$$H_{c20} = 0.68 T_c \left(\frac{dH_{c2}}{dT}\right)_{Tc} \qquad (3)$$

Equations (1) and (2) taken together in the form

$$\rho = \frac{3.27 \times 10^{-5}}{\gamma}\left(\frac{H_{c20}}{T_c}\right) = \frac{2.23 \times 10^{-5}}{\gamma}\left(\frac{-dH_{c2}}{dT}\right)_{T_c} \qquad (4)$$

show that for single band superconductors both the slope $H_{c20}/T_c$ and $(-dH_{c2}/dT)_{Tc}$ scale with $\rho$, and hence that (for fixed $T_c$) both the low temperature $H_{c2}$ and that near $T_c$ benefit simultaneously from an increase in $\rho$. Some treatments of the subject prefer to treat $\rho$ in terms of an electron diffusivity, $D$ ($= v_F^2 \tau/3$, where $\tau$ is the impurity-scattering relaxation time) such that $1/\rho = e^2 N_F D$, in which case the scaled benefits to $H_{c20}/T_c$ and $(-H_{c2}/dT)T_c$ are seen to vary as $1/D$.

Equation (3) suggests an experimental method of obtaining $H_{c20}$ (which may be outside the range of many laboratory magnets) that can be carried out at moderate fields, near $T_c$. Equations (1)–(4) indicate that across-the-board increases in $H_{c2}$ can be expected to follow increases in impurity resistivity, $\rho$, – a rule which has helped to guide the design of low temperature superconductors over the years. But whereas it is qualitatively true also for MgB$_2$ the simple rule has quantitative variants, for example: (i) some MgB$_2$ samples with comparable $H_{c2}$s have been found to have very different $\rho$s – an Eqn. (1) departure. (ii) Some experimentally measured H$_{c20}$s have been found to be considerably greater than the $(dH_{c2}/dT)_{Tc}$-predicted values – an Eqn (3) departure.

## *Upper Critical Fields of MgB$_2$ -- General Responses to Changes in $D_\sigma$ and $D_\pi$*

Although Eqn. (3) predicts an $H_{c20}$ some 68% below a linear back extrapolation of the $H_{c2}(T)$ vs $T$ line near $T_c$, measurements on some MgB$_2$ samples have yielded $H_{c20}$ values that lie close to and even above that extrapolation. With reference to a copious body of experimental results, Gurevich et al [21], based on the detailed analysis of Gurevich [22], have been able to further illustrate and explain this behavior in terms of the electronic diffusivities $D_\pi$ and $D_\sigma$ associated with MgB$_2$'s two-band conductivity:

$$\sigma = e^2 (N_{F\pi} D_\pi + N_{F\sigma} D_\sigma) \qquad (5)$$

As explained by the above authors, the introduction of two diffusivities results in pronounced departures from the single-band predictions. In particular $(H_{c2}/T_c)$ and $(-dH_{c2}/dT)_{Tc}$ respond individually to $D_\pi$ and $D_\sigma$ rather than together and in proportion to $1/D$. Thus, according to Gurevich et al [21] and to a first approximation:

$$H_{c2nearTc} \propto \frac{1}{\lambda_1 D_\sigma + \lambda_2 D_\pi} \qquad (6)$$

where the $\lambda$s are the coupling constants and

$$H_{c20} \propto \frac{1}{\sqrt{D_\sigma D_\pi}} \qquad (7)$$

Simply stated, near $T_c$ $H_{c2}$ varies inversely as the weighted arithmetic mean of $D_\sigma$ and $D_\pi$, while $H_{c20}$ varies inversely as their geometric mean. Thus while increasing impurity scattering is beneficial to $H_{c2}$ over the entire temperature range for both single-band and two-band superconductors, in the latter case (and recognizing the independence of $D_\pi$ and $D_\sigma$) it offers $H_{c20}$ the opportunity to diverge to very large values -- well beyond the 68% of the $(-dH_{c2}/dT)_{Tc}$ extrapolation -- in response to strong decreases in either $D_\sigma$ or $D_\pi$.

## Detailed Responses to $D_\sigma$, $D_\pi$ Inequalities

Suppose, as has turned out to be the case for dirty MgB$_2$ films, that $D_\pi << D_\sigma$ (i.e $\pi$ scattering much stronger than $\sigma$ scattering) then Eqn. (6) approximates to

$$H_{c2nearTc} \propto \frac{1}{\lambda_1 D\sigma} \qquad (8)$$

In general $H_{c2\ near\ Tc}$ is controlled by the greater of the two diffusivities in this case $D_\sigma$. Further changes in $H_{c2\ near\ Tc}$ would expect to follow a manipulation of $D_\sigma$. On the other hand, since in Eqn. (7) $D_\sigma$ and $D_\pi$ are still tied together as a product (even though $D_\pi$ is the "dominant" component), manipulation of either of them would seem to influence $H_{c20}$. But examination of the $H_{c20}$ relationship in more detail reveals that for very different diffusivities

$$H_{c20} \propto \frac{1}{D_\sigma} \qquad D_\sigma << D_\pi \qquad (9)$$

and

$$H_{c20} \propto \frac{1}{D_\pi} \qquad D_\pi << D_\sigma \qquad (10)$$

So $H_{c20}$ is controlled by the **smaller** of the two diffusivities – the $D_\pi$ in dirty MgB$_2$ films.

## $H_{c2}$ Anisotropy

We conclude with a note on the effects of scattering on changes in $H_{c2}$ anisotropy. Equations (6) and (7) refer to the field-perpendicular (to the basal plane, *a-b*) orientation. For the field-parallel orientation (*c*-direction), the diffusivities must be transformed, for example by $D_\sigma = \sqrt{(D_\sigma^{ab} D_\sigma^c)}$ and the same for $D_\pi$. Let us consider the anisotropy parameter $\gamma = H_{c2||}/H_{c2\perp}$ at low temperatures and high temperatures on the assumption that $D_\pi << D_\sigma$, as in the case of dirty MgB$_2$ film. A temperature dependent $H_{c2}$ anisotropy arises via the differing anisotropies of $D_\sigma$ and $D_\pi$ coupled with the fact that in this case $D_\pi$ controls $H_{c2o}$ while $D_\sigma$ controls $H_{c2}$ near $T_c$. It can be shown (and is to be expected intuitively) that the diffusivity in the 3-D $\pi$ band is less anisotropic than that in the 2-D $\sigma$ band and thus $\gamma$ decreases as the temperature decreases. Thus the enhancement of $H_{c20}$ and the softening of its anisotropy is a consequence of the two-band character of MgB$_2$ coupled with a $D_\pi << D_\sigma$.

## Experimental

### Sample preparation

A CTFF process was used to produce MgB$_2$/Fe composite strand [10,11]. The starting Mg powders were 325 mesh 99.9% pure, and the B powders were amorphous, 99.9% pure, and at a typical size of 1–2 µm. The powders were V-mixed and then run in a planetary mill; in some cases nano-size SiC powders were added. After powder preparation, a tube mill was used to continuously dispense the powder onto a strip of high purity Fe. This, after closing, was inserted into a monel or Cu-30Ni tube. Heat treatments were then performed under Ar. Ramp up times were typically x h, and the samples were furnace cooled. The times and temperatures at the plateau ranged from 675-850°C, for times between 5-30 minutes.

*Measurements*

Four-point transport $J_c$ measurements were made on two sample types, short samples and barrel samples. The short samples were 3 cm in length, with a gauge length of 5 mm). The barrels had a single layer of wire, with each turn separated from its neighbors. The ends were soldered onto Cu end-rings, and the current leads were attached over a 1 m segment of strand. The voltage taps were 50 cm apart. Standard Pb–Sn solder was used for forming the contacts, and the $J_c$ criterion was 1 µV/cm. Most measurements were made at 4.2 K in background fields of up to 15 T (applied transverse to the strand). In a few cases, vibrating sample magnetization (VSM) measurements were made. The VSM system had a 9 T maximum field, and loop measurement times were typically 20 min in duration. Measurements were performed at 4.2 K.

**Results**

Transport and magnetic $J_c$s were measured at 4.2 K for a set of samples (Set I, see Table 1) with, and in one case without, SiC additions. The transport results were measured on barrel samples in this case (1 m segments of wire). These transport results were then converted to $F_p$, using $F_p = J_cB$, with the results displayed in Figure 1. Magnetic results were measured on short sections, and gave somewhat higher results. The most likely reason for this is the presence of inhomogenieties along the length of the strand, resulting in an somewhat (artificially) suppressed $F_p$ for the transport results. While this does display reasonable transport properties over 1 m lengths, the influence of SiC additions are unclear. This led is to a new set of experiments, Set II. In order to limit the scatter generated by extrinsic defects, it was decided to measure short samples. However, transport measurements were chosen over magnetic, to insure that any SiC were present in the wires in a usable way.

Set II consisted of six strands, six with SiC, and six without (see Table 1). The HT schedules were similar, with temperatures of 700°C and 800°C, and times from 5 to 30 min. These strands were then measured for transport $J_c$ at 4.2 K in fields of up to 15 T. The results are displayed in Figure 2 in the form of a Kramer plot. The superiority of the SiC samples is immediately clear. Linear extrapolations for all samples show a jump in the irreversibility field of between 1-2 T. Low current data was taken for samples SiC800/05 and NSC700/15 which show a high field (low current) deviation from linearity, however, these curves show the same jump in $H_r$. Clearly, SiC is increasing $H_r$, although any dependence of this effect on HT is unclear. Figure 3 shows this data in the form of an $F_p$ curve. $F_p$, as well as $H_r$, is apparently increased by the SiC, at least near 4.2 K.

.     The last set of data was given a very short HT at an elevated temperature; about 5 minutes at 900°C. $J_c$ for this sample is shown in Figure 4 for a variety of temperatures. However, we can immediately notice that $H_r$ is at least 16 T – significantly higher than for Set II samples. This suggests that the higher temperature HT is (perhaps with the addition of SiC) enhancing $H_r$. Figure 5 shows a Kramer plot for this data, and we notice the high temperature tail of the plot, at 4.2 K, leads off to an $H_r$ of about 18 T. $F_p$ curves, now normalized to $H_r$, are presented in Figure 6. We see by comparison to Figure 3 that $F_{pmax}$ seems to be enhanced. $F_{pmax}$ for Figure 3 samples, even with SiC, are going to be les than 10 $GN/m^3$, while that for Figure 6 data is 20 $GN/m^3$. Not only that, but the field at which the maximum occurs is higher – in this case to more that 4 T – higher than that of Figure 3

**Discussion and Conclusions**

SiC clearly increases $H_r$ at 4.2 K for the present samples. This is in agreement with the results of [16] where enhancements in high field $J_c$ are seen, and furthermore $c$-axis and $a$-axis lattice parameters are modified by the inclusion of SiC. Additionally, an increase in $F_p$ is seen. This is consistent with the TEM images of [16], which show an increased dislocation density. We also note that the saturation of the change in $c$-axis and $a$-axis  parameters with SiC addition seen in [16] suggest a limit to the SiC influence. Looking to results of the same group on carbon nano-tube additions, this may be controlled by reaction temperature [17]. This would explain the significant improvements in $H_r$ and $F_p$ for higher reaction temperatures seen in Figures 4-6.
In summary, in-situ processed strands with 10% SiC additions HT at 700-800C show improved $H_r$ and $F_p$ values as compared to control samples, with $H_r$ increasing by 1.5 T. At 900°C even larger improvements are seen, with $H_r$ reaching 18 T and $F_p$ values maximizing at 20 $GN/m^3$. High field measurements are now needed, but it may be that SiC increases both $B_{c2}$ and pinning.

**Acknowledgements**

This work was supported by a State of Ohio Technology Action Fund Grant.

| Name | Tracer ID | SiC? | HT | SC% | OD |
|---|---|---|---|---|---|
| **Set I** | | | | | |
| BNSC700/30 | | N | 700/30 | 22.8 | 0.984 |
| BSiC700/05 | | Y | 700/5 | 24.6 | 0.836 |
| MSiC700/5 | | Y | 700/5 | 24.6 | 0.836 |
| MSiC700/15 | | Y | 700/15 | 24.6 | 0.836 |
| **Set II** | | | | | |
| SiC700/30 | 256a | Y | 700/30 | 27.5 | 0.827 |
| SiC800/15 | 256a1 | Y | 800/15 | 27.5 | 0.827 |
| SiC800/05 | 256a2 | Y | 800/05 | 27.5 | 0.827 |
| NSC700/30 | 253b | N | 700/30 | 24.1 | 0.791 |
| NSC800/15 | 253b1 | N | 800/15 | 24.1 | 0.791 |
| NSC700/15 | 253b2 | N | 700/15 | 24.1 | 0.791 |
| **Set III** | | | | | |
| D | 257 | Y | 700/5+900/5 | XXX | XXX |

Table 1. Strand Specifications

**Fig. 1**

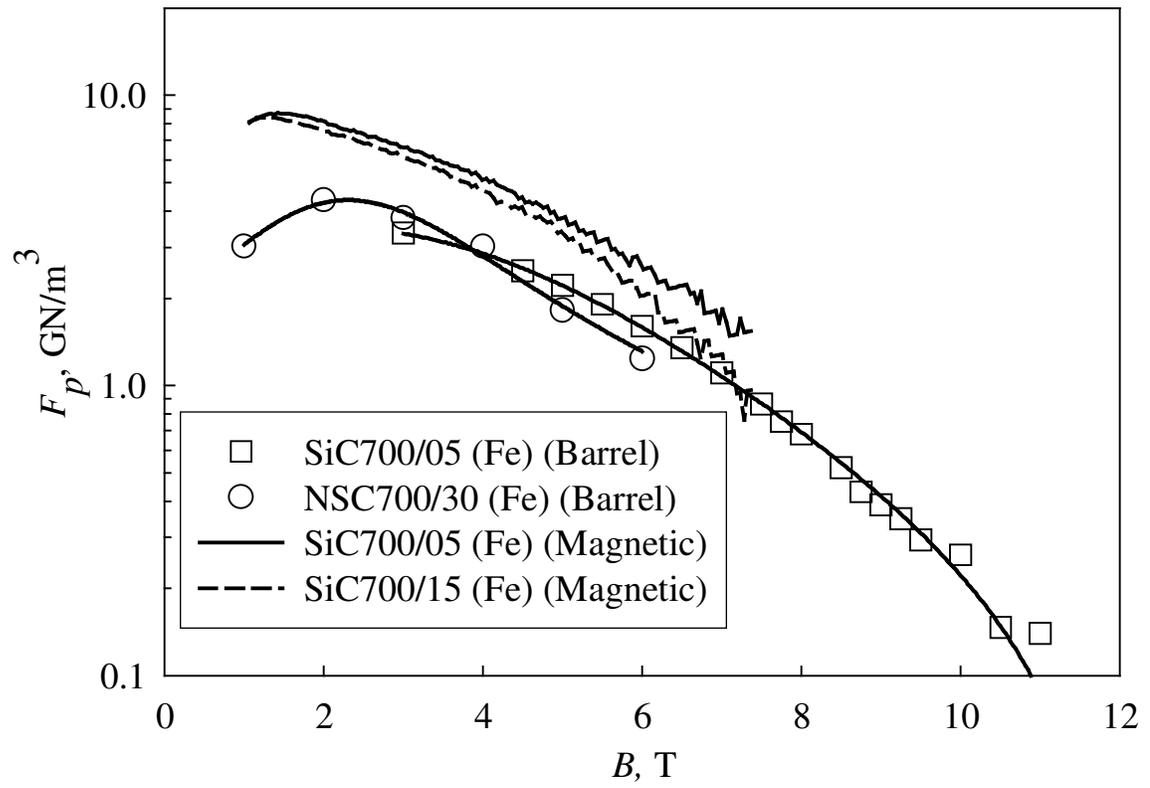

**Fig. 2**

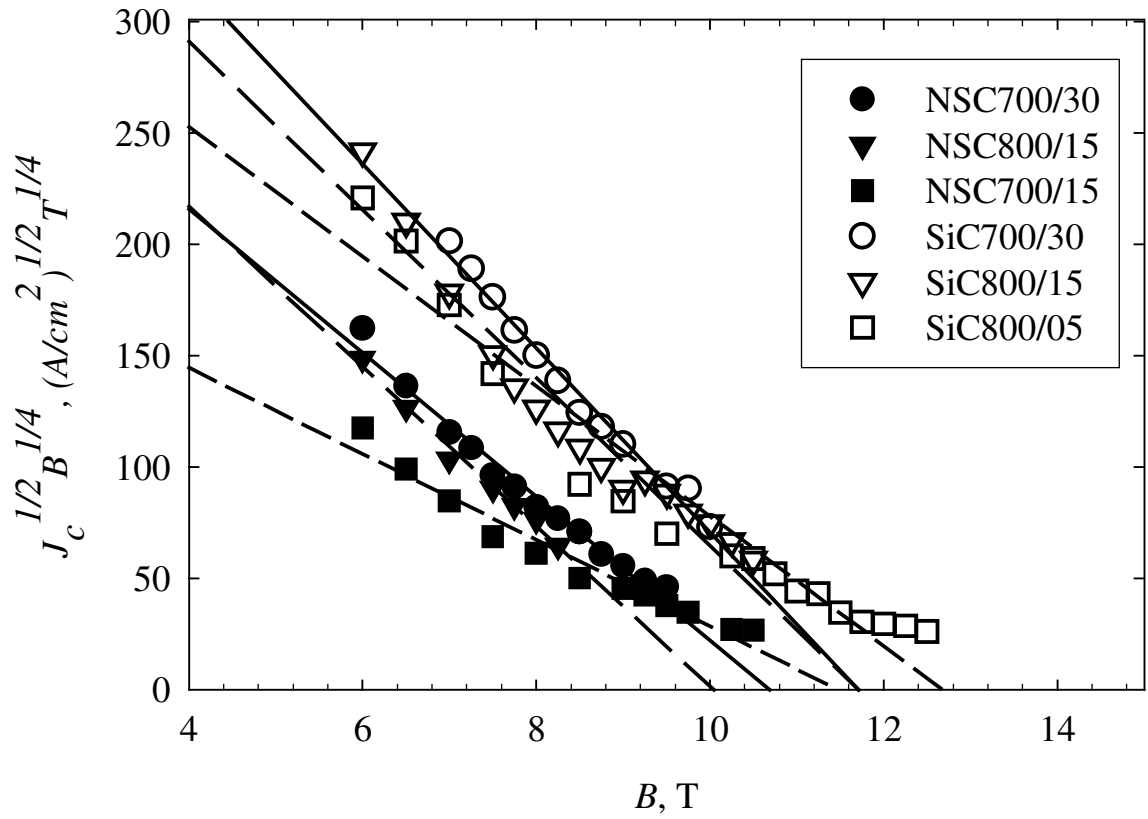

**Fig. 3.**

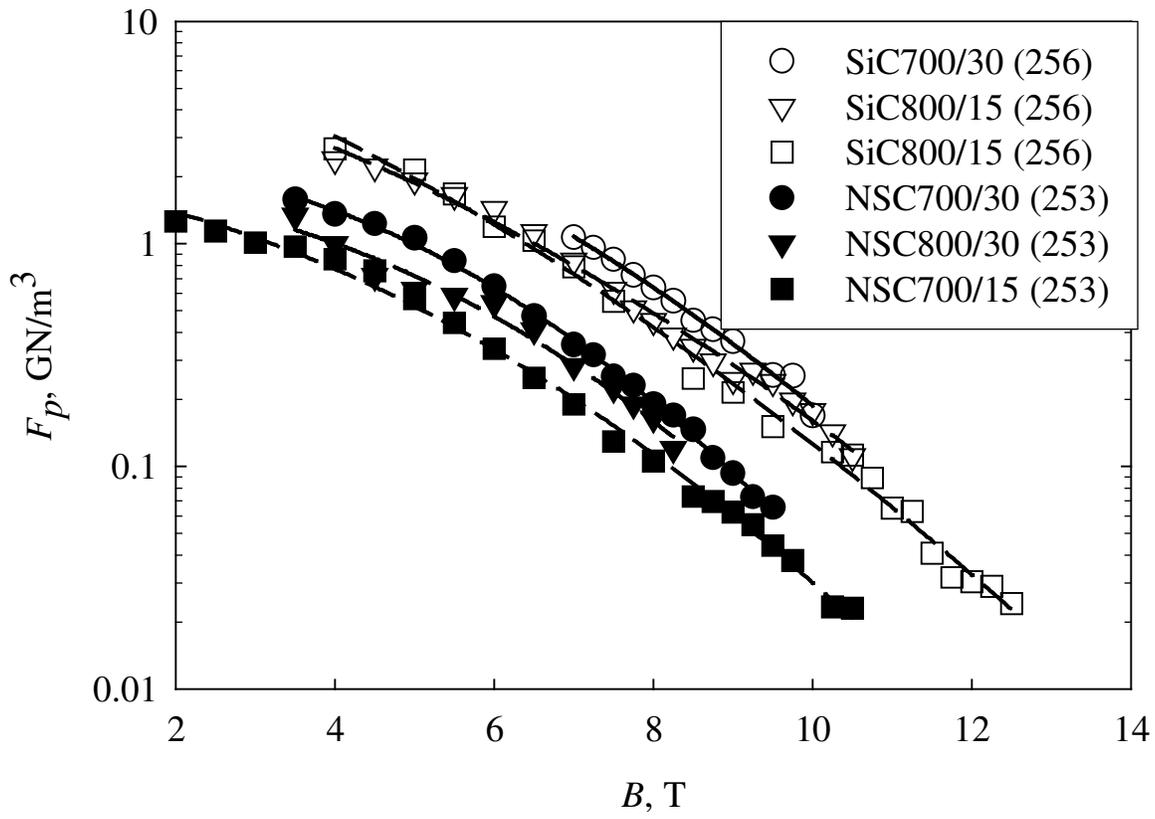

**Fig. 4.**

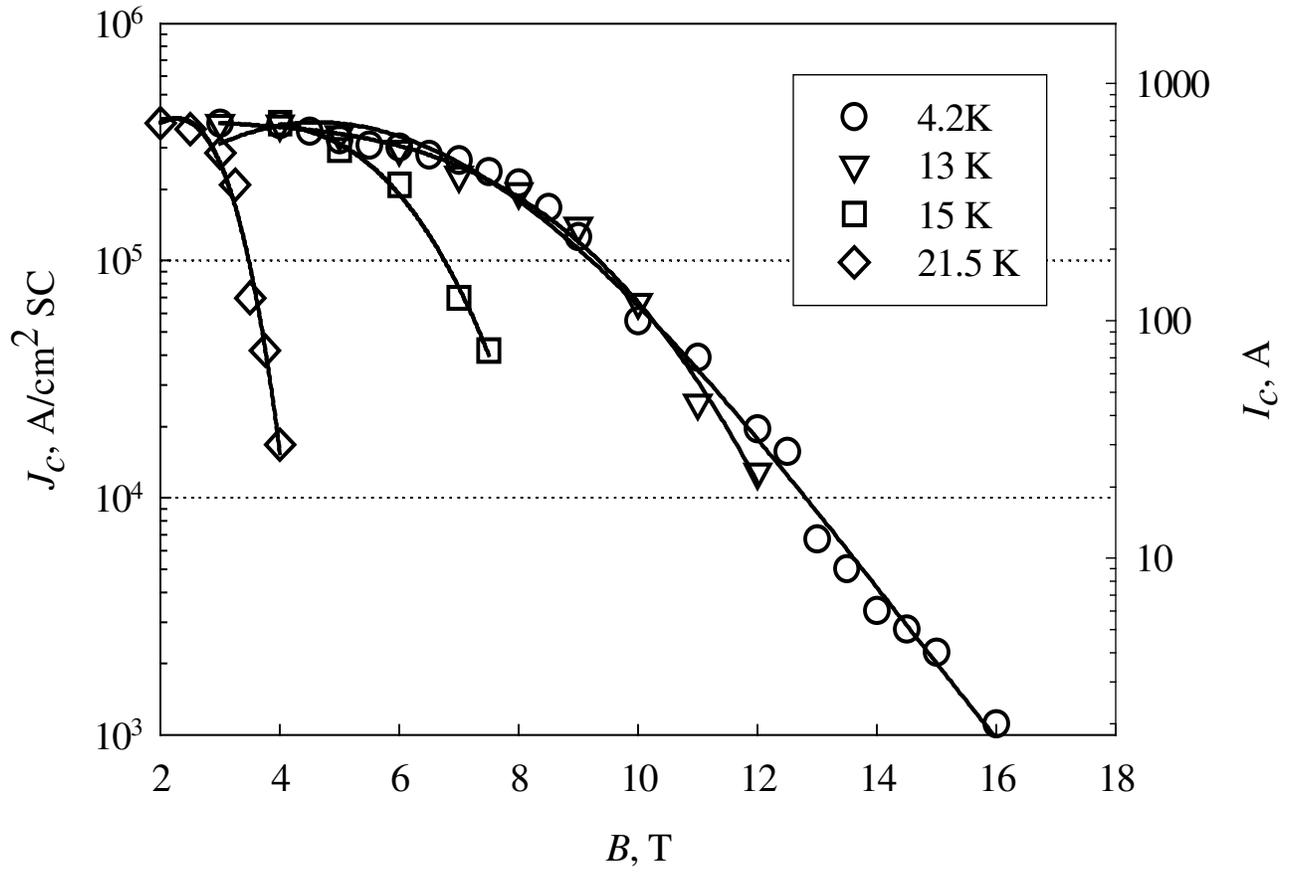

**Fig. 5.**

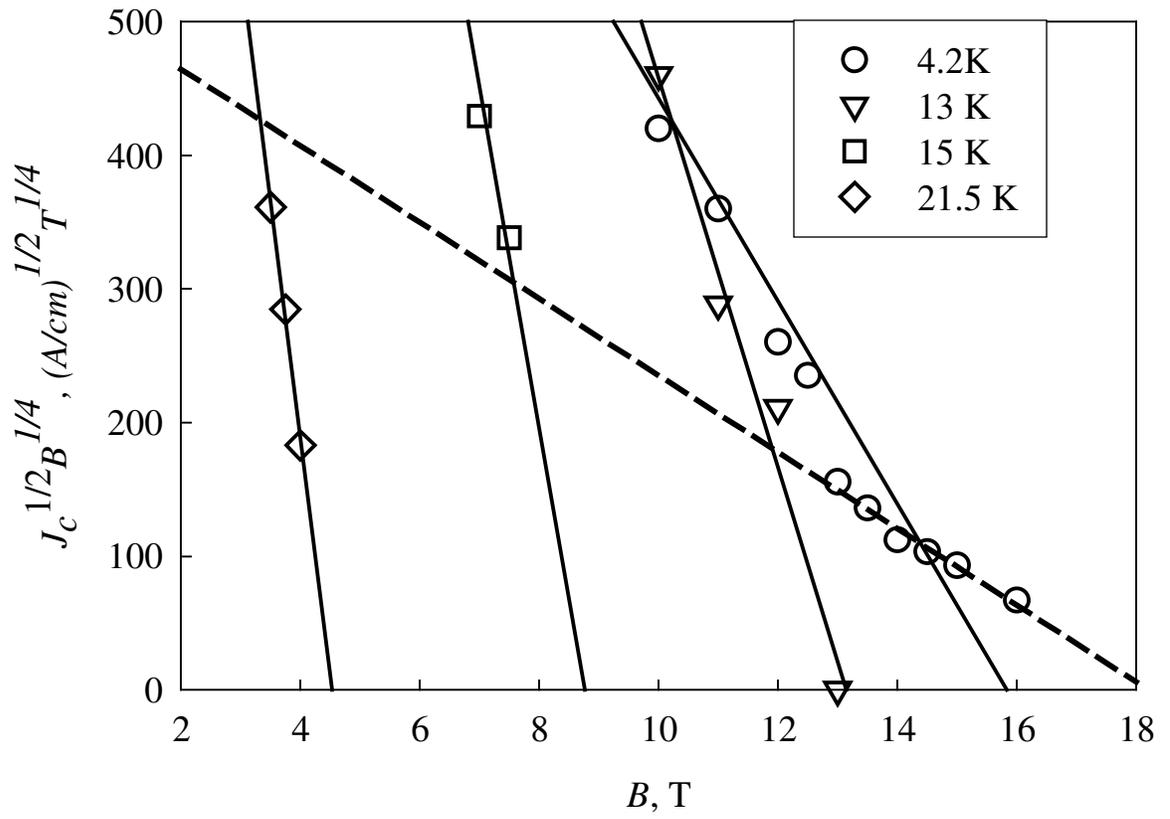

**Fig. 6.**

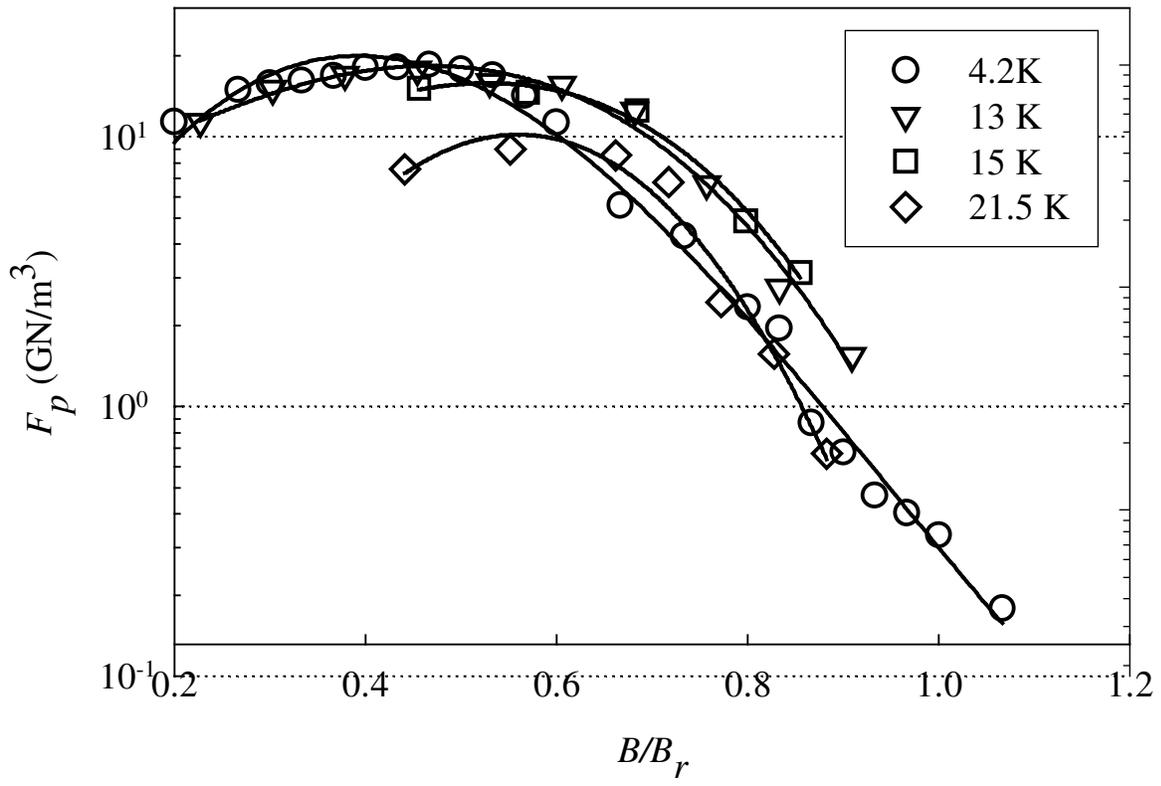